\documentclass[%
reprint,
superscriptaddress,
showpacs,showkeys,
 amsmath,amssymb,
 aps,
floatfix,
]{revtex4-1}

\usepackage{graphicx,subfigure}
\graphicspath{{Figures/}}
\usepackage{multirow}

\newcounter{reaction} 
\begin{document}
\bibliographystyle{revtex4}
\draft


\title[HiPIMS Ni$_{80}$Fe$_{20}$]{Comparison of magnetic and structural properties of permalloy Ni$_{80}$Fe$_{20}$
grown by dc and high power impulse magnetron sputtering}



\author{Movaffaq Kateb}
\affiliation{Science Institute, University of Iceland,
Dunhaga 3, IS-107 Reykjavik, Iceland}

\author{Hamidreza Hajihoseini}
\affiliation{Science Institute, University of Iceland,
Dunhaga 3, IS-107 Reykjavik, Iceland}
\affiliation{Department of Space and Plasma Physics, School of Electrical Engineering and Computer Science, KTH--Royal Institute of Technology, SE-100 44, Stockholm, Sweden}

\author{Jon Tomas Gudmundsson}
\affiliation{Science Institute, University of Iceland,
Dunhaga 3, IS-107 Reykjavik, Iceland}
\affiliation{Department of Space and Plasma Physics, School of Electrical Engineering and Computer Science, KTH--Royal Institute of Technology, SE-100 44, Stockholm, Sweden}

\author{Snorri Ingvarsson}
\email[Corresponding author email address:]{sthi@hi.is}
\affiliation{Science Institute, University of Iceland,
Dunhaga 3, IS-107 Reykjavik, Iceland}
\date{\today}

\begin{abstract}

We study the microstructure and magnetic properties of Ni$_{80}$Fe$_{20}$ thin films grown by high power impulse magnetron sputtering (HiPIMS), and compare with films grown by dc magnetron sputtering (dcMS). The films were grown under a tilt angle of 35$^{\circ}$ to identical thickness of 37~nm using both techniques, at different pressure ($0.13-0.73$~Pa) and substrate temperature (room temperature and 100~$^{\circ}$C). All of our films display effective in-plane uniaxial anisotropy with square easy axis and linear hard axis magnetization traces.  X-ray diffraction reveals that there is very little change in grain size within the pressure and temperature ranges explored. However, variations in film density, obtained by X-ray reflectivity measurements, with pressure have a significant effect on magnetic properties such as anisotropy field ($H_{\rm k}$) and coercivity ($H_{\rm c}$). Depositions where adatom energy is high produce dense films, while low adatom energy results in void-rich films with higher $H_{\rm k}$ and $H_{\rm c}$. The latter applies to our dcMS deposited films at room temperature and high pressure. However, the HiPIMS deposition method gives higher adatom energy than the dcMS and results in dense films with low $H_{\rm k}$ and $H_{\rm c}$. The surface roughness is found to increase with increased pressure, in all cases, however it showed negligible contribution to the increase in $H_{\rm k}$ and $H_{\rm c}$.

\end{abstract}

\maketitle


\section{Introduction}

Permalloy Ni$_{80}$Fe$_{20}$, referred to as Py hereafter, is a very well known ferromagnetic
material and has over the years been used extensively in various industrial applications. In
its thin film form it presents a (surprisingly) low in-plane easy-axis-like anisotropy caused
by competing contributions from Ni and Fe. A further benefit is its vanishing magnetostriciton.
It has been employed in many applications, such as anisotropic magnetoresistance (AMR) and
planar Hall effect (PHE) field sensors \cite{McGuire1975}, magnetic recording heads
\cite{andricacos98:671,craus05:363,cooper05:103,yokoshima10:67} and magnetoresistive random
access memory (MRAM) \cite{akerman05:508,wang17:072401}.  Thin permalloy films have been
deposited by thermal evaporation \cite{umezaki82:753,sugita1967}, electroplating
\cite{shimokawa12:2907,theis13:853,yanai18:056123}, electron beam deposition
\cite{watanabe05:2390}, rf diode sputtering \cite{collins81:165,yang89:3734,potzlberger84:851},
dc magnetron sputtering (dcMS) \cite{potzlberger84:851,rijks97:362,svalov2010} and rf magnetron
sputtering \cite{chaug90:505}. It is well known that the electrical and magnetic properties are
influenced by the deposition conditions and method \cite{collins81:165,yang89:3734}. We have
recently shown that a tilt deposition geometry with respect to the substrate normal using dcMS
induces strong in-plane uniaxial anisotropy, i.e.\ a square easy axis with sharp transitions
and a linear hard axis without hysteresis \cite{Kateb2017,Kateb2018}. Those are desirable
properties for both magnetic memories and field sensors. However, tilt sputtering at high
pressures, which is more appropriate for large industrial applications, has not been studied
yet. This is mainly because tilt deposition requires low pressure as it suffers from scattering
of the sputtered flux by the working gas at high pressures \cite{barranco2016}. There is also a
competition between in-plane and out-of-plane anisotropy in Py films. It has been shown that
for both dcMS and evaporation, increased pressure reduces in-plane uniaxial anisotropy
accompanied with losing magnetic softness \cite{sugita1967,fujiwara1968,svalov2010}. Regardless
of the deposition method, the poor in-plane anisotropy was associated with formation of stripe
domains at thicknesses beyond the critical thickness. The critical thickness decreases
dramatically with increased pressure \cite{sugita1967,fujiwara1968,svalov2010}, e.g.\ it is
about 250~nm at 0.38~Pa and decreases to 60~nm at 2.4~Pa using dcMS deposition
\cite{svalov2010}. Since, higher pressure increases surface roughness and encourages void-rich
structure in the case of Py \cite{Kools1995}, a decrease in the critical thickness was
attributed to an increase in the surface roughness \cite{svalov2010} and increase in defects
and voids \cite{sugita1967,fujiwara1968} with increased pressure.  

An interesting solution to overcome void-rich structure and rough surface is offered by high
power impulse magnetron sputtering (HiPIMS) which is an ionized physical vapor deposition
technique that has attracted much interest lately \cite{helmersson06:1,gudmundsson12:030801}.
By pulsing the target to a high power density with unipolar voltage pulses, low duty cycle, and
low repetition frequency, high electron density is achieved in the plasma
\cite{helmersson06:1,gudmundsson12:030801}. This high electron density leads to a high
ionization fraction of the sputtered material. As a result HiPIMS presents denser
\cite{samuelsson10:591,magnus11:1621,magnus12:1045}, void free \cite{alami05:278} and smoother
coatings \cite{alami05:278,magnus11:1621,hajihoseini2017} compared to conventional sputtering methods. It has been shown that amorphous magnetic films of Fe$_{73.5}$CuNb$_3$Si$_{15.3}$B$_7$ can be grown by HiPIMS and they are claimed to have the same composition as the target over wide range of pressures
\cite{velicu12:1336}. This is important since dcMS has been found to present 2.3~$at.\%$ change in iron content of Py by changing the pressure in the 0.38 -- 2.4~Pa range \cite{svalov2010} which
can have significant effect on the magnetic properties, as discussed e.g.\ by \citet[p.~190 \&
369]{ohandley2000}. It has also been shown that the coercivity ($H_{\rm c}$) of the films
grown by HiPIMS increases with increased pressure in the 1.33 -- 8.00~Pa range
\cite{velicu13:1329}. However, in later studies \cite{velicu12:1336,velicu13:1329} the same
authors compared magnetic softness of various films grown at different pressures but with
\emph{different thickness}. They also did not explore the variation of surface roughness and
film density at different pressures and their effect on the magnetic properties. Thus the
effect of pressure change on the microstructure as well as magnetic properties of films grown
by HiPIMS are not well understood, and in particular the role of film density and surface
roughness.

Here, we study the properties of Py films grown by HiPIMS at different pressures while
maintaining the same thickness and compare with dcMS grown films under similar conditions. The
main focus is on studying deposition at increased pressure and finding under which conditions
we are able to maintain high quality magnetic films, with well defined square hysteresis loops
and low coercivity and anisotropy field using dcMS and HiPIMS deposition, respectively.

\section{Experimental apparatus and method}

The substrates were square 10$\times$10~mm$^2$ p-Si(001) with a native oxide of about 2.4~nm thickness. The Py thin films were grown in a custom built UHV magnetron sputtering chamber with a base pressure less than $5 \times 10^{-7}$~Pa. The deposition was done with argon of 99.999~\% purity as the working gas using a Ni$_{80}$Fe$_{20}$ at.\% target with 75~mm diameter. The substrate was kept at room temperature (21$\pm$0.1~$^\circ$C) and 100~$^\circ$C during growth, respectively.

In order to ensure as uniform film thickness as possible, we rotate the sample 360$^\circ$ in
one direction and then stop (due to electrical wiring to the sample holder) and then rotate it
back by 360$^\circ$ in the reverse direction. The rotation is at $\sim$12.8~rpm with 300~ms
stop time at the turning points. The deposition is done under an angle of 35$^\circ$ with
respect to the substrate, the stop-and-turn position plays an important role in defining
magnetization axis direction \cite{Kateb2017}. In short, tilt angle induces hard magnetization
axis along the plane of incidence and easy magnetization axis perpendicular to that. We have
already shown that the tilt deposition can have a stronger effect on magnetization direction
than the applied magnetic field during growth. Further details on sample growth and a schematic
of our deposition geometry can be found in reference \cite{Kateb2017}.  

The dcMS depositions were performed at four specific pressures in the range of 0.13 -- 0.73~Pa at 150~W dc power (MDX 500 power supply from Advanced Energy). For HiPIMS deposition, the power was supplied by a SPIK1000A pulse unit (Melec GmbH) operating in the unipolar negative mode at constant voltage, which in turn was charged by a dc power supply (ADL GS30). The discharge current and voltage were monitored using a combined current transformer and a voltage divider unit (Melec GmbH). The pulse length was 250~$\mu$s and the pulse repetition frequency was 100~Hz. At all pressures, the average power during HiPIMS deposition was maintained around 153~W, to be comparable with dcMS grown films at 150~W. The HiPIMS deposition parameters were recorded by a LabVIEW program communicating with our setup through high speed data acquisition (National Instruments).

X-ray diffractometry (XRD) was carried out using a X'pert PRO PANalitical diffractometer (Cu K$_\alpha$ line, wavelength 0.15406~nm) mounted with a hybrid monochromator/mirror on the incident side and a 0.27$^\circ$ collimator on the diffracted side. A line focus was used with a beam width of approximately 1~mm. Grazing incidence (GI)XRD scans were carried out with the incident beam at $\theta = 1^\circ$. The film thickness, density and surface roughness was determined by low-angle X-ray reflectivity (XRR) measurements with an angular resolution of 0.005$^\circ$. The film thickness, density and roughness were obtained by fitting the XRR data using the commercial X'pert reflectivity program, that is based on the Parrat formalism \citep{parratt54:359} for reflectivity.

Magnetic hysteresis was characterized using a home-made high sensitivity magneto-optical Kerr effect (MOKE) looper. The coercivity is read directly from the easy axis loops. The anisotropy field is obtained by extrapolating the linear low field trace along the hard axis direction to the saturation magnetization level, a method commonly used when dealing with effective easy axis anisotropy.

\section{Results and discussion}
We deposited Py thin films using HiPIMS and dcMS. We discuss the discharge characteristics of HiPIMS in section \ref{discharge}, the films' microstructure in section \ref{microstructure} and characterize their magnetic properties in section \ref{magnetic-properties}. 
\subsection{Discharge current and voltage waveforms}
\label{discharge}
\begin{figure}[h]
	\includegraphics[width=1\linewidth]{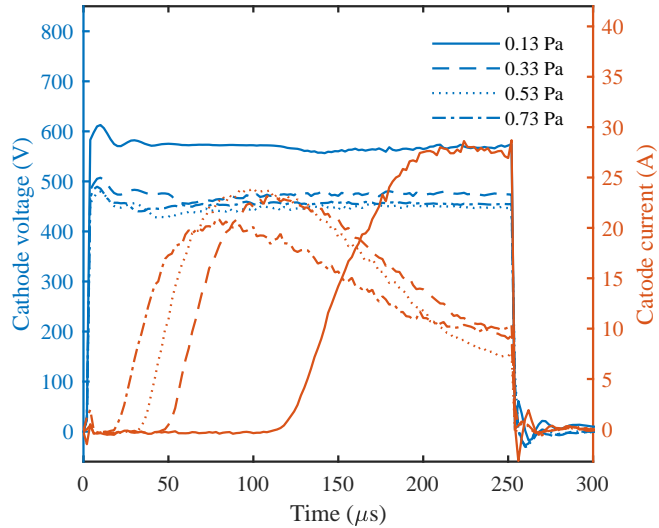}
	\caption{\label{waveform} The discharge current-voltage waveform at different pressures, for a 75~mm diameter Ni$_{80}$Fe$_{20}$ target with Ar as working gas. The line style of the legend applies to both discharge voltage and current traces.}
\end{figure}

Figure \ref{waveform} shows the current and voltage waveforms of the HiPIMS discharge recorded during room temperature growth at different pressures. It can be seen that a nearly rectangular voltage pulse of 250~$\mu$s length was applied to the cathode target. Beside oscillations at the beginning and after ending the voltage pulse, there are local minima due to the initial current rise in all cases. The oscillations are due to an internal inductance of the power supply which creates a resonance circuit with the parasitic capacitance of cables and the capacitance of the cathode target. Since the current onset occurs at different times for different pressure, the value of applied voltage (height in the blue curves) changes to maintain the required average power ($\sim$153~W here). Table \ref{dcV} summarizes cathode voltages used in our depositions which indicate that HiPIMS voltage pulses are well above dcMS discharge voltages.
\begin{table}[h]
\centering
\caption{Summary of cathode voltage in our depositions at different pressures and applied power. Voltage values are in V.}
\label{dcV}
\begin{tabular}{c|c c c c}
	\hline
    \hline
	 method and & \multicolumn {4}{|c}{Pressure (Pa)}\\
	 average power & 1.3 & 3.3 & 5.3 & 7.3 \\
	\hline
    HiPIMS 150~W & 583 & 489 & 460 & 459 \\
	dcMS 150~W & 406 & 322 & 300 & 295 \\
	dcMS 50~W & 370 & 306 & 288 & -- \\
    \hline\hline
\end{tabular}
\end{table} 

\begin{figure}[h]
	\includegraphics[width=1\linewidth]{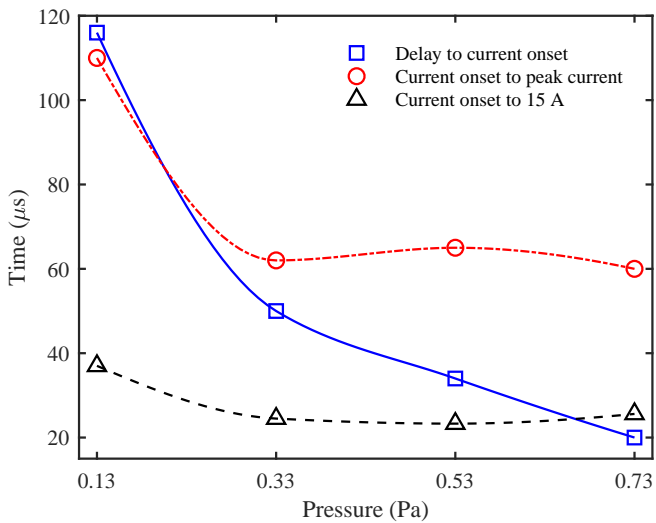}
    \caption{\label{time} The variation of delay time to current onset and time for current to rise from onset to peak current and 15~A versus Ar pressure.}
\end{figure}

The current waveforms can be described by three distinct regions, as previously described by Lundin \emph{et al.} \cite{lundin09:045008} (I) plasma initiation and a current maximum, followed by (II) a decay to a minimum and then (III) a steady state regime that remains as long as the discharge voltage level is maintained. The initial peak current is a result of strong gas compression due to the rapid large flux of sputtered atoms coming from the target. Within a few $\mu$s collisions of the sputtered atoms with the working gas atoms leads to heating and expansion of the working gas, known as rarefaction. As a result, the working gas atoms are replaced by the sputtered atoms in the vicinity of the cathode target to some extent as the pulse evolves. However, it has been shown that the rarefaction is primarily due to ionization losses in the target vicinity \cite{huo12:045004}. The rarefaction causes the discharge current to fall as can be seen for pressures in the range 0.33 -- 0.73~Pa.
In this regard, the 0.13~Pa case is different than for higher pressures in all three stages. This is due to a long delay on the current initiation which results in appearance of only the first stage of the discharge current during the voltage pulse. At higher pressures, the second stage is also observable, while the pulse length is not long enough to see the third stage of the current evolution. Figure \ref{time} shows variation of the delay time and time between current initiation to peak current and 15~A as a function of Ar pressure. The current initiation delay time changes linearly in the 0.33 -- 0.73~Pa range and increases dramatically at lower pressures. Previously we have reported the increased delay time with decreasing Ar pressure when sputtering a tantalum target \cite{gudmundsson02:249} and for a vanadium target in Ar/N$_2$  mixture \cite{hajihoseini2017}. Due to its stochastic nature, the delay time can be explained statistically as described by Yushkov and Anders \cite{yushkov10:3028}. In a simplified way, the probability of ionization depends on availability of precursor, mainly represented by the pressure of the working gas and ratio of applied voltage to voltage required for ionization of the gas and sputtered atoms. Thus, Yushkov and Anders \cite{yushkov10:3028} model the inverse of delay time to be proportional to the cathode voltage. They also proposed a linear variation of delay time with inverse of pressure at constant cathode voltage. However, here we preferred to maintain constant power by increasing the cathode voltage as pressure decreases. Thus there is a competition between cathode voltage increment and pressure decrement to shorten and lengthen the delay time, respectively. Since the delay became longer as the pressure decreased, it can be concluded that in the present study the pressure has a dominant effect over cathode voltage on the length of delay time.
 
The time required to reach the peak current after current initiation is 115~$\mu$s at 0.13~Pa while for higher pressures it stands nearly constant at about 75~$\mu$s. Since the peak currents are not equal at different pressures, we also calculate rise time of 0 to 15~A at each pressure. The graph shows the rate of current rise is 0.41~A/$\mu$s at 0.13~Pa and it increases to 0.61~A/$\mu$s by increasing the pressure to 0.33~Pa. At higher pressures the current rise rate remains almost unchanged.

\subsection{Microstructure}
\label{microstructure}

\subsubsection{X-ray reflectivity}
\begin{figure}[h]
\includegraphics[width=1\linewidth]{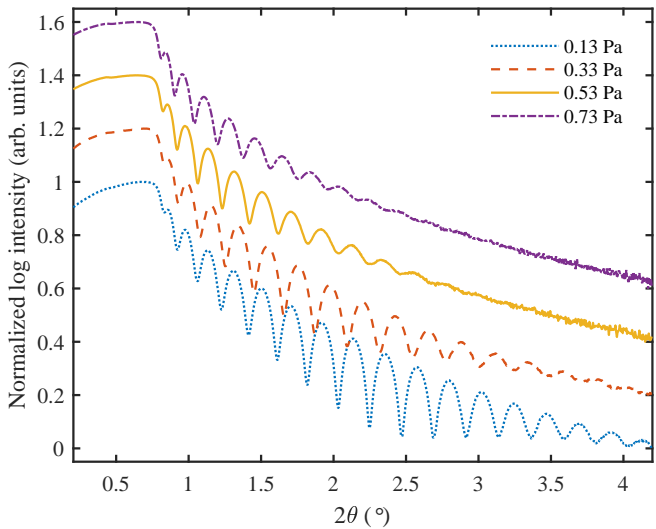}
\caption{\label{XRR} The XRR curves for the film grown with dcMS at room temperature and at various pressures up to nearly the same thickness. The figures are offset vertically for illustration purposes.}
\end{figure}

Figure \ref{XRR} shows the X-ray reflectivity curves of the films grown with dcMS at room temperature as an example. The figure clearly shows a change in X-ray reflectivity with pressure. At higher pressures, the amplitude of the oscillations decays faster with incident angle, which represents greater surface roughness of the film \cite{Li2015}. This behavior is reproduced similarly for the rest of the pressure series grown at different conditions but are not shown here. In order to obtain the most precise estimates of thickness, density and surface roughness of the films, the reflectivity curves have to be fitted carefully. To this end we take into account formation of an oxide layer and adsorbed moisture on its surface which is reasonable since the measurements were performed in ambient atmosphere (\emph{ex-situ}).

\begin{figure}[h]
	\includegraphics[width=1\linewidth]{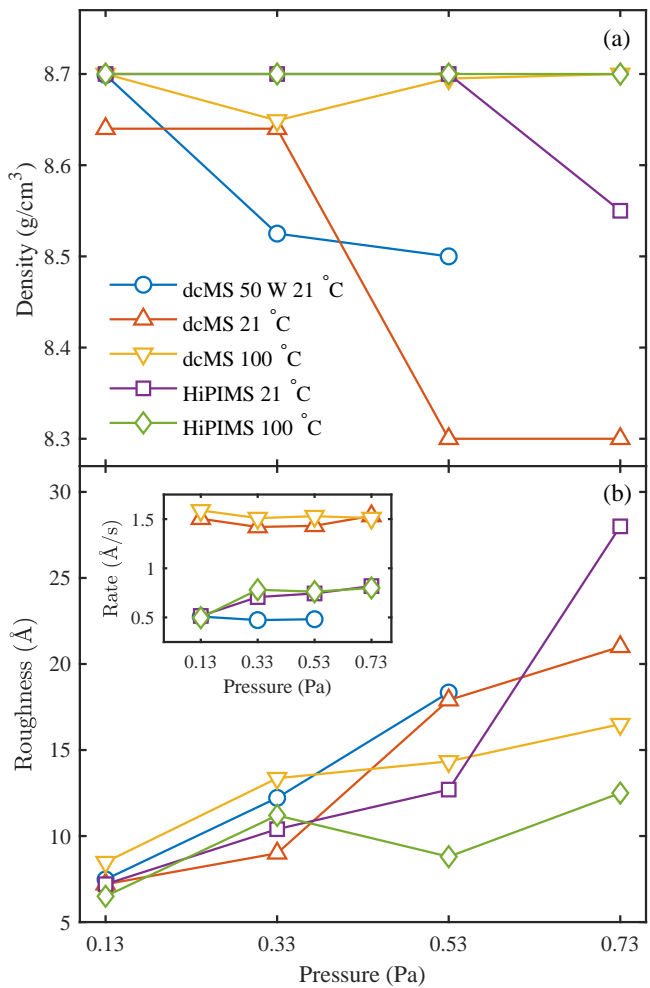}
	\caption{\label{XRRfit} The (a) film density and (b) surface roughness of the films obtained by fitting to XRR curves for the film grown at different pressure with nearly the same thickness. The figure inset shows deposition rate.}
\end{figure}

The results of the fitting are shown in figure \ref{XRRfit} for both of the deposition methods
and both substrate temperatures. All the films were grown to the same thickness of 37~nm and
the deposition rate is shown in the figure inset. We note that HiPIMS deposition has
significantly lower deposition rate than dcMS deposition. As shown in figure \ref{XRRfit}(a),
the density of the films grown by dcMS at room temperature shows an abrupt drop in the pressure
range between 0.33 -- 0.53~Pa. In contrast, utilizing HiPIMS at room temperature can maintain
high density for most of the cases explored. The only exception occurs at 0.73~Pa which density
shows deviation from the almost constant high density attained in other HiPIMS deposited samples. One may
think that since the average deposition rate of HiPIMS is significantly lower than for dcMS
(cf.\ figure \ref{XRRfit}(b) inset) it may result in higher densities. Thus another series was
grown by dcMS at 50~W power which gives deposition rate of 0.5~{\AA}/s equal to the average
deposition rate of HiPIMS at 0.13~Pa. This lower deposition rate at 50~W dcMS improves the
density at 0.13~Pa, compared to the 150~W counterpart, but the film density drops as the
pressure is increased. Thus the low deposition rate is not solely responsible for the high film
density obtained by HiPIMS at room temperature. For both dcMS and HiPIMS growth, increased
substrate temperature seems to efficiently maintain the film density at a value very close to
the bulk density of 8.72~g/cm$^3$ \cite[p.~548]{ohandley2000}. 
0.33~Pa which shows slight deviation from maximum density.  This can be explained by the fact
that at higher pressures the mean free path is reduced and an adatom experiences more
collisions and loses more kinetic energy before arriving at the surface. However, raising the
substrate temperature to increase the adatom mobility at the surface maintains high density of
the film during growth at higher pressures. 

The surface roughnesses of the films grown at different conditions are shown in figure \ref{XRRfit}(b). It is worth noting that the different growth conditions have the minimum effect at the lowest pressure, 0.13~Pa, while the values are more scattered at higher pressures. In general, the trend in surface roughness is an increase with increasing pressure. Our results show precisely such a trend. Also it should be noted that films grown at $100~^\circ$C with both deposition methods show less roughness than their lower temperature counterparts grown at higher pressures. Again, this is in agreement with the statement that the more opportunity an adatom has to seek a desirable site on the surface the smoother and denser the resulting film.

For the dcMS grown films, the deposition rate is nearly independent of the pressure variation
but with slightly higher values at 100~$^\circ$C. However, the HiPIMS growth at 0.13~Pa
presents considerably lower average deposition rate at both room temperature and 100~$^\circ$C.
This is due to longer delay time at 0.13~Pa which is evident from the trends shown in figures
\ref{waveform} and \ref{time}. We would like to remark that the average deposition rate of
HiPIMS can be somewhat misleading i.e.\ by accounting for 100~Hz pulses of 250~$\mu$s
length and neglecting the delay time, the effective deposition time is below 25~ms per second which
gives an effective value of $\sim$40 times the average deposition rate.

We would like to remark that utilizing a weaker magnet in the magnetron might change the behavior of the deposition rate in both dcMS and HiPIMS. For instance, we have noticed that reducing the magnet's field strength to half the value and doubling the target thickness would result in less confinement of plasma with a linear reduction of deposition rate with pressure increment (not shown here). The linear decrease in deposition rate with increased pressure in the 1.33 -- 8~Pa range was reported using HiPIMS \cite{velicu13:1329}. But the main focus of the current study is on the effect of pressure and substrate temperature thus we preferred a constant deposition rate to reduce the number of contributing parameters. The above mentioned effect of the magnetic field strength on HiPIMS deposition was demonstrated earlier and the interested reader is referred to our earlier work on VN deposition \cite{hajihoseini2017} for further information.

\subsubsection{X-ray diffraction}
\begin{figure}[h]
	\includegraphics[width=1\linewidth]{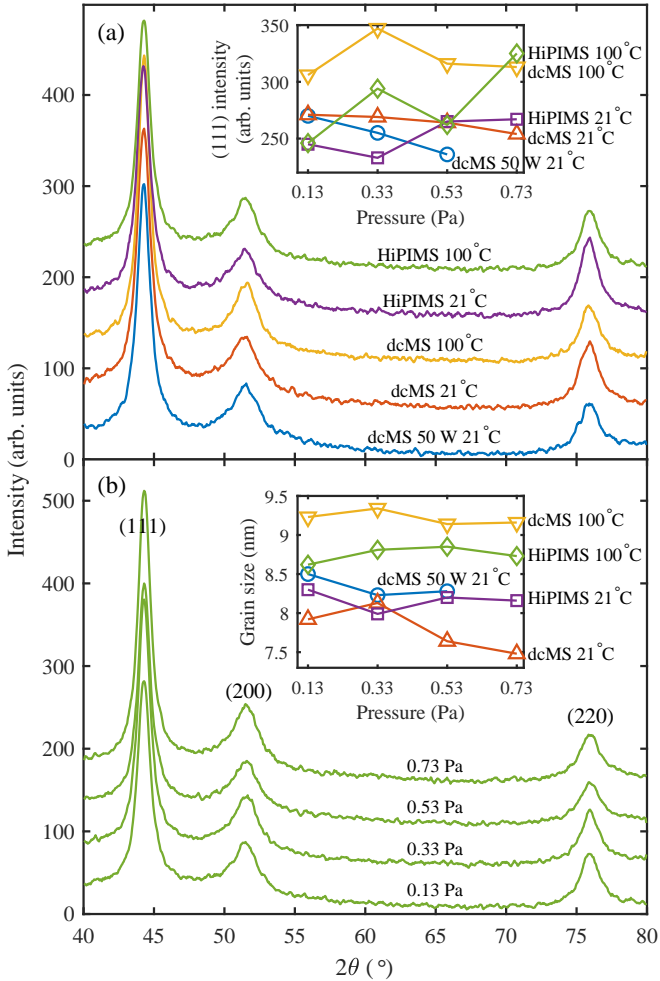}
	\caption{\label{GIXRD} The GIXRD patterns of (a) different films grown at 0.13~Pa (b) films grown at different pressures using HiPIMS at 100$^\circ$C. The figure insets shows (111) peak intensity and grain size estimated from the (111) peaks using the Scherrer equation.}
\end{figure}

Figure \ref{GIXRD} shows two sets of GIXRD patterns, namely different films grown at fixed pressure of 0.13~Pa with dcMS and HiPIMS at varying temperature and for films grown by HiPIMS at 100$^\circ$C at different pressures. Results for other films are not shown here since they show very little difference. The figure insets depict the variation in the (111) peak intensity and estimated grain size from the Scherrer equation \citep{langford1978}, which has been proven to give quantitatively correct values for Py films \citep{neerinck1996}, with pressure for all of the films. The three main peaks are evident in all cases those are located at 44.217, 51.518 and 75.845$^\circ$ corresponding to (111), (200) and (220) planes, respectively \cite{dzhumaliev2016}. The dominant peak is (111) in all cases. The (111) texture provides perpendicular anisotropy to
the Py films as for fcc alloy structures the $\langle111\rangle$ direction is the easy
magnetization axis \citep[p.~224]{ohandley2000}. This becomes important for films somewhat
thicker than ours and plays an important role in stripe domain formation.  It has been shown
for films grown by normal deposition geometry, that an increase in the pressure reduces the
peak height and this is most pronounced for the (111) peak \cite{svalov2010}. Only our dcMS
grown films at room temperature are in agreement with those results. In contrast, maximum (111)
peak intensity is obtained at 0.33~Pa for dcMS grown films at 100~$^\circ$C, and at 0.73~Pa for
the films grown by HiPIMS at both room temperature and 100~$^\circ$C. The (200) peak intensity
shows a slight increase while (220) peak intensity presents a decrease with increasing pressure
for all the films (not shown here).

A more quantitative understanding of the film crystallinity is represented by the grain size. The estimated grain size from the (111) peaks shows negligible variation with pressure. It is also worth noting that dcMS growth results in smaller grain size at room temperature than HiPIMS while it gives larger grain size at 100~$^\circ$C. Thus increasing the substrate temperature has a more pronounced effect in the grain growth during dcMS deposition.

\subsection{Magnetic properties}
\label{magnetic-properties}
\subsubsection{Anisotropy and coercive fields}
\begin{figure}[h]
\includegraphics[width=1\linewidth]{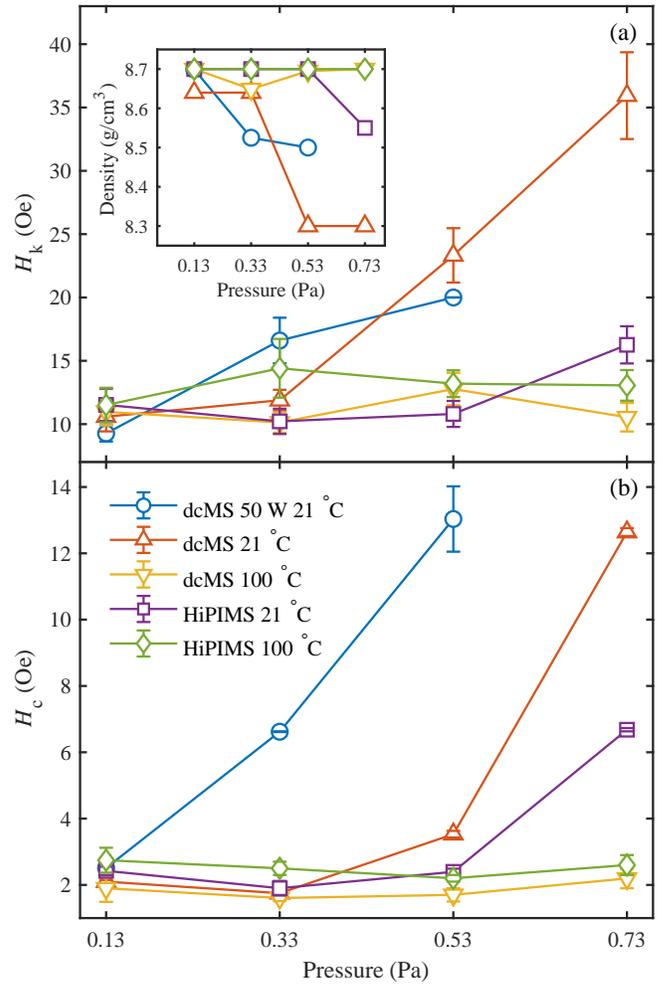}
\caption{\label{MOKE} The (a) anisotropy field $H_{\rm k}$ and (b) coercivity $H_{\rm c}$ of the films grown to the same thickness at different pressures. These values are extracted from MOKE measurements along hard and easy axis, respectively. The figure inset shows density variation at different pressures.}
\end{figure}

Figure \ref{MOKE} shows the variation of the anisotropy field $H_{\rm k}$ and coercivity $H_{\rm c}$ with pressure for both dcMS and HiPIMS grown films. The results are extracted from hysteresis loops measured along easy and hard axis of the film using MOKE. It can be clearly seen that $H_{\rm k}$ increases with increased pressure for films grown by dcMS at room temperature. All our other films have either a nearly constant $H_{\rm k}$ (dcMS and HiPIMS at 100~$^\circ$C) or a delayed and slower growth as function of pressure as in the case of the HiPIMS grown film at room temperature. Both increased substrate temperature and the higher ion energy involved in HiPIMS deposition contribute to more adatom surface diffusion, encouraging defect-free crystal growth that helps to maintain the low anisotropy field $H_{\rm k}$. 

It has been shown previously, that associated with void formation there is an increase in the anisotropy field $H_{\rm k}$ in Py \cite{sugita1967,fujiwara1968}. Our quantitative results are in good agreement with their interpretation. The density results shown in the inset in figure 6 (a) show that the density remains high for the
films that exhibit low $H_{\rm k}$ but drops in the cases where there is an increase in $H_{\rm k}$. Presumably the lower density is associated with more defective crystal growth and void formation. 
Considering the fact that surface roughness increases with pressure in all cases (cf.~figure
\ref{XRRfit}(b)), our results are inconsistent with the results of Choe and Steinback
\cite{choe1999} who found a linear reduction in $H_{\rm k}$ with increased surface roughness. A
reason for this difference might be the film thickness i.e.\ their films were 15~nm thick which
makes them more sensitive to the surface properties compared to 37~nm films here which are more
influenced by ``bulk-like'' properties. Thus, knowing that the grain size is nearly constant
for each pressure series (as shown in figure \ref{GIXRD} inset) the variation of $H_{\rm k}$
here is associated with the density of the films. The high film density is maintained in 0.13~Pa films using HiPIMS and also dcMS at 100~$^\circ$C while the $H_{\rm k}$ is low.

The variation of $H_{\rm c}$ with pressure is shown in figure \ref{MOKE}(b). It is clear that
at a certain threshold pressure there is a density reduction accompanied by an increase in
$H_{\rm c}$ for films grown at room temperature.  For the dcMS films this threshold pressure is
at 0.13 and 0.53~Pa, respectively for 50 and 150~W power, while for the room temperature HiPIMS
sample it is at 0.73~Pa.  Again due to higher film density for growth at 100~$^\circ$C, the low
$H_{\rm c}$ is maintained at high pressures. Based on surface roughness Choe and Steinback
\cite{choe1999} reported two regions: (I) below 8~{\AA} surface roughness where $H_{\rm c}$
slightly increases with surface roughness and (II) surface roughnesses higher than 8~{\AA}
where $H_{\rm c}$ increases dramatically with surface roughness, due to surface roughness
induced pinning of domain walls during magnetization reversal. This is not the case in our
results. For instance HiPIMS grown film at 21~$^\circ$C and 0.73~Pa present highest roughness
of the all films while it presents an intermediate $H_{\rm c}$ compared to the dcMS
counterpart. In the present results, the pinning is attributed to voids and defects appearing
in films with low mass density.

\subsubsection{Magnetization traces}

\begin{figure}[h]
\includegraphics[width=1\linewidth]{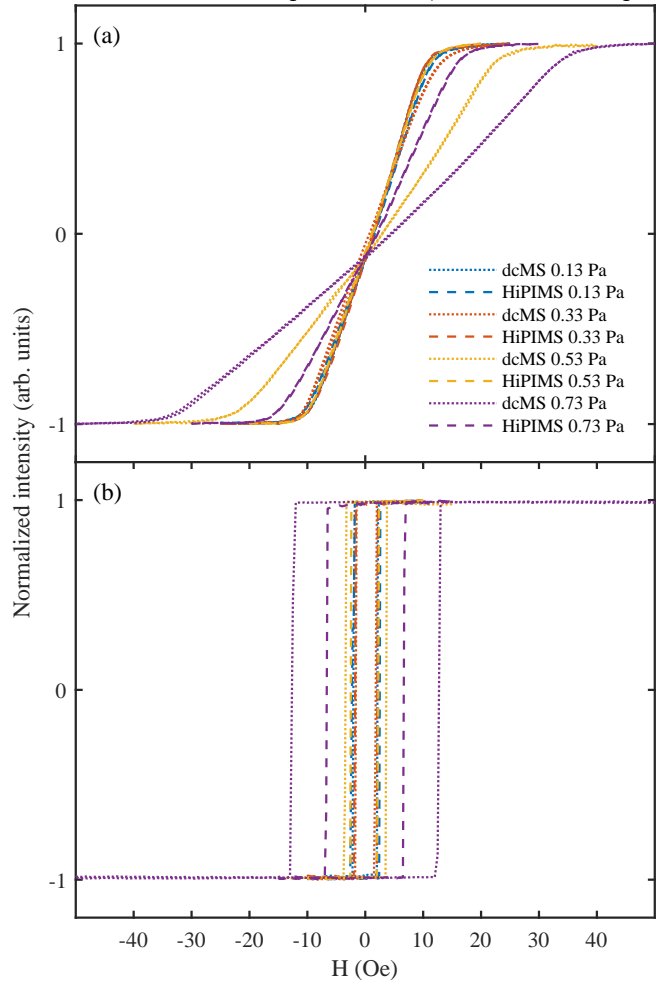}
\caption{\label{loops} The MOKE response of the films grown by dcMS and HiPIMS at room temperature measured along (a) hard and (b) easy axis of the films.}
\end{figure}

In addition to desirable values of $H_{\rm k}$ and $H_{\rm c}$, for many applications it is also important to have well defined magnetic axes. Figure \ref{loops} shows the average of a few loops obtained by MOKE measurements for room temperature grown films. Similar results were obtained for films grown at 100~$^\circ$C, that are not shown here. The dotted lines belong to dcMS grown films which always present higher saturation fields than their HiPIMS counterparts grown at the same pressure, shown by dashed lines. This is more evident at higher pressures of 0.53 and 0.73~Pa shown in yellow and purple, respectively.

It can be seen that in both methods, deposition under an angle provides a linear non-hysteretic
hard axis and a square easy axis with sharp transitions. However, it has been shown previously
that dcMS deposition normal to substrate at high pressure with \emph{in-situ} magnetic field,
gives open hard axis and rounded easy axis traces \cite{svalov2010}. We have shown previously
that at low pressures (0.13~Pa) the effect of deposition under an angle on the direction of the
magnetization axis can be stronger than the effect of an \emph{in-situ} magnetic field during
growth \cite{Kateb2017,Kateb2018}. The present results show that uniaxial anisotropy induced by
tilt deposition maintains the shape of loops at different pressures. Although we have detected
an increase in both $H_{\rm k}$ and $H_{\rm c}$ in some of the samples grown at room
temperature (cf.\ figure \ref{MOKE}) there is no indication of out-of-plane magnetization or
poorly defined uniaxial anisotropy in those samples.

\section{Conclusions}

We have demonstrated the deposition of Ni$_{80}$Fe$_{20}$ thin films using high power impulse magnetron sputtering, HiPIMS. For comparison we also deposited films using dc magnetron sputtering under the same conditions, i.e.\ to the same thickness, at the same pressure, substrate temperature, tilt angle and with power identical to the HiPIMS average power. We compared the results of structural characterization (X-ray) and magnetic properties. The results indicate that the higher the adatom energy, as it meets with the sample substrate/film, the denser the film, accompanied with low coercive and anisotropy fields. All conditions kept the same, the
HiPIMS deposition method gives a higher adatom energy than dcMS. Increased adatom energy can also be achieved by raising the deposition temperature or lowering the pressure. In accordance with
this our results show a drop in film density for samples deposited at room temperature (our lowest deposition temperature) with increasing pressure, accompanied by a rise in both coercive and anisotropy
fields.

\acknowledgements

This work was partially supported by the University of Iceland Research Funds for Doctoral students, the Icelandic Research Fund Grants Nos.~130029 and 120002023, and the Swedish Government Agency for Innovation Systems (VINNOVA) contract no. 2014-04876.

\bibliography{heim78}
\bibliographystyle{apsrev}

%
\end{document}